# Optimal Voltage and Current Control of an HVDC System to Improve Real-Time Frequency Regulation

Do-Hoon Kwon, *Member, IEEE* and Young-Jin Kim, *Member, IEEE*

*Abstract*—High-voltage direct-current (HVDC) systems for constant or intermittent power delivery have recently been developed further to support grid frequency regulation (GFR). This paper proposes a new control strategy for a line-commutated converter-based (LCC) HVDC system, wherein the DC-link voltage and current are optimally regulated to improve real-time GFR in both rectifier- and inverter-side AC networks. A dynamic model of an LCC HVDC system is developed using the DC voltage and current as input variables, and is integrated with feedback loops for inertia emulation and droop control. A linear quadratic Gaussian (LQG) controller is also designed for optimal secondary frequency control, while mitigating conflict between the droop controllers of the HVDC converters. An eigenvalue analysis is then conducted, focusing on the effects of model parameters and controller gains on the proposed strategy. Simulation case studies are also performed using the Jeju-Haenam HVDC system as a test bed. The results of the case study confirm that the proposed strategy enables the HVDC system to improve GFR, in coordination with generators in both-side grids, by exploiting the fast dynamics of HVDC converters. The proposed strategy is also effective under various conditions for the LQG weighting coefficients, inertia emulation, and droop control.

## I. INTRODUCTION

ADVANCES in power conversion and control technologies have brought about a renewed interest and resurgence in the applications of high-voltage direct-current (HVDC) systems for economical, reliable inter-regional delivery of electric power. In particular, several HVDC projects worldwide aim to facilitate the connection of large-scale renewable energy sources (RESs) in remote host regions with distant load centers. For example, a line-commutated converter-based (LCC) HVDC system installed in the Midwest region of the United States has been considered to deliver approximately 25% of the power generated by the wind turbines, with a total capacity of 16 GW, to the Eastern Interconnection [1], [2]. It has also been suggested that new control strategies should be devised for LCC HVDC systems to support real-time grid frequency regulation (GFR), alleviating the impact of intermittent wind power on both sending- and receiving-end grids.

The real-time GFR of LCC HVDC systems has been widely studied. For example, in [3], an HVDC rectifier station was connected to an offshore wind farm (WF). The rectifier firing angle or, equivalently, DC-link current, was controlled for GFR in the rectifier-side grid. In [4], an HVDC system was used as an interface between a WF and an inverter-side network, so that its rotational kinetic energy could be exploited in the form of electrical inertia to improve GFR. The paper [5] analyzed the effects of a short circuit ratio and a DC inertia constant on GFR in a weak grid, which was linked to both WF and LCC HVDC systems. In [6], a fuzzy inference system was integrated into the damping controller of an HVDC inverter to suppress oscillations of the inverter-side frequency and AC voltage, resulting from a short-circuit fault at the point of common coupling. However, in [3]–[6], an LCC HVDC system was controlled to support GFR only in either the inverter- or rectifier-side network, because the other-side network was assumed to include only WFs, rather than synchronous generators and loads.

Recently, GFR has been considered on both sides of the LCC HVDC system, for example, in [7]–[10]. The hourly optimal power flow was calculated in [7] considering the fast-acting, corrective control of an LCC HVDC system when the HVDC bi-pole block resulted in large frequency deviations in the sending- and receiving-end grids. In [8], a four-state nonlinear model was presented to determine the interaction between the rectifier- and inverter-side grids and develop HVDC firing angle controllers and a WF droop controller. Considering inter-grid coupling, optimal control schemes can also be adopted to better utilize HVDC systems. For example, in [9] and [10], linear quadratic regulators (LQRs) were used for optimal centralized and decentralized control, respectively, of HVDC systems, to enhance the damping of inter-area power oscillations; however, the HVDC system models were relatively simple and droop controllers were not considered.

In [7]–[9], the DC-link voltage of the HVDC system was simply maintained at the rated value. In practice, LCC HVDC converters can operate under normal conditions, with short-term under- and over-voltages of the DC link, as reported in field test results [11], [12]. Recently, a few studies have been carried out on time-varying control of the DC-link voltage, mainly for a type of voltage-source converter-based HVDC system. For example, in [13], the HVDC system was controlled with the operating range of the DC-link voltage varying from approximately 0.98 pu to 1.04 pu for a 5% variation in the load demand. In [14], the energy stored in the DC link was combined with the frequency support capabilities of wind turbines. Time-varying control of the DC-link voltage can also be applied to LCC HVDC systems. For example, in [15] and [16], the DC-link voltage control was discussed to improve the capabilities of LCC HVDC systems for short-term power transfer and reactive power control, respectively.

This paper proposes a new control strategy for an LCC

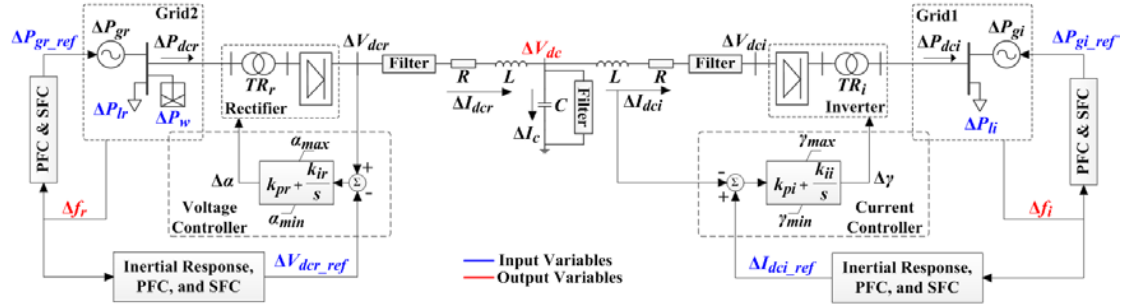

Fig. 1. A schematic diagram of the proposed control strategy for an LCC HVDC system to support real-time GFR in the rectifier- and inverter-side AC networks.

HVDC system, in which the optimal control of DC-link voltage and current is achieved in real time to improve GFR in both the rectifier- and inverter-side networks. A small-signal model of the LCC HVDC system is developed using the DC voltage and current references as input variables, and validated via comparison with a comprehensive model of a real HVDC system [15], [17]. The HVDC converters support real-time GFR via primary (PFC) and secondary frequency control (SFC). For PFC, the converters are integrated with feedback loops to emulate inertial response, and frequency- and DC voltage-power droop control. A linear quadratic Gaussian (LQG) controller, combining an LQR with a Kalman filter, is also incorporated into the feedback loops to achieve optimal SFC, minimizing the weighted sum of instantaneous and accumulated deviations of the DC-link voltage and grid frequencies in the rectifier- and inverter-side networks. An eigenvalue analysis is then conducted to evaluate the performance of the proposed control strategy with variation in the model parameters and controller gains of the HVDC system. Simulation case studies are also carried out to demonstrate that the proposed strategy is effective for improving real-time GFR with variation in load demands and WF power generation under various conditions for the LQG weighting factors and the inertia emulation and droop control approaches.

## II. Dynamic Model of LCC HVDC System for GFR via the Control of DC-link Voltage and Current

Fig. 1 shows a schematic diagram of the proposed strategy, in which an LCC HVDC system supports real-time GFR in the rectifier- and inverter-side networks. The HVDC converters are equipped with inner feedback loops to control $\Delta V_{dcr}$ and $\Delta I_{dci}$. The DC link is characterized using a T-model, where $\Delta P_{dcr}$ flows into the rectifier and $\Delta P_{dci}$ flows from the inverter. At both HVDC terminals, second-order DC filters are installed to mitigate the effects of DC-side harmonic currents [18], [19]. Moreover, the WF is installed on the rectifier-side grid, causing frequency deviations, $\Delta f_r$ and $\Delta f_i$, on both sides.

### A. Dynamic Model of LCC HVDC System

The DC-link voltage and current at the terminals of the rectifier and inverter can be represented [20] as:

$$V_{dcx} = \frac{3\sqrt{2}N}{\pi TR_x} V_{lx} \cos\beta - \frac{3X_{cx}N}{\pi} I_{dcx}, \quad (1)$$

$$\frac{\sqrt{2} X_{cx} TR_x I_{dcx}}{V_{lx}} = \cos\beta - \cos(\beta + \mu_x), \quad (2)$$

where descriptions of the variables are provided in Section IV.

Note that the subscript $x$ is replaced with $r$ and $i$ to indicate the variables related to the rectifier and inverter, respectively. Moreover, $\beta$ corresponds to $\alpha$ and $\gamma$ for the firing and extinction angles, respectively. Then, $P_{dcr}$ and $P_{dci}$ are represented as:

$$P_{dcx} = V_{dcx} I_{dcx}, \quad (3)$$

$$= \frac{3N}{4\pi X_{cx}} \left(\frac{V_{lx}}{TR_x}\right)^2 \{\cos(2\beta) - \cos 2(\beta + \mu_x)\}. \quad (4)$$

Furthermore, Fig. 1 shows that $\Delta\alpha$ and $\Delta\gamma$ can be estimated as:

$$\Delta\beta = (k_{px} + k_{ix}/s)(\Delta H_{dcx} - \Delta H_{dcx\_ref}), \quad (5)$$

where $H$ corresponds to $V$ and $I$ for the voltage and current controllers, respectively: i.e., $(x, \beta, H) = (r, \alpha, V)$ or $(i, \gamma, I)$. From (1)–(3) and (5), $\Delta P_{dci}$ can be obtained as:

$$\Delta P_{dci} = (V_{dci0} - a_{2i}I_{dci0})\Delta I_{dci} - a_{3i}b_{1i}\left(k_{pi} + \frac{k_{ii}}{s}\right)(\Delta I_{dci} - \Delta I_{dci\_ref}), \quad (6)$$

where the coefficients $a$ and $b$ are provided in Appendix A. By linearizing and combining (4) with (5), $\Delta P_{dci}$ can also be expressed as:

$$b_{3i}\Delta P_{dci} = a_{4i}b_{5i}\Delta I_{dci} + a_{3i}(b_{2i}b_{5i} - b_{3i}b_{4i})\left(k_{pi} + \frac{k_{ii}}{s}\right)(\Delta I_{dci} - \Delta I_{dci\_ref}). \quad (7)$$

From (6) and (7), the transient response $\Delta I_{dci}/\Delta I_{dci\_ref}$ of the inverter can be represented as (8): see the bottom of the page where the coefficients $c$ are also presented in Appendix A. For real HVDC systems, $\Delta I_{dci}/\Delta I_{dci\_ref}$ in (8) can be approximated to a strictly proper first-order transfer function, because the ratio of $m_3$ to $m_1$ has a magnitude comparable to that of $V_{dcr0}$ to $I_{dcr0}$: e.g., $V_{dcr0}$ = 184,000 V, $I_{dcr0}$ = 407.6 A, and $m_3/m_1$ = 7.7×10$^{-3}$ for a real 150-MW HVDC system [21]. Similarly, from the linearized set of (1)–(5), the transient response of the rectifier to the reference voltage can be expressed as:

$$\frac{\Delta V_{dcr}}{\Delta V_{dcr\_ref}} = \frac{a_{3r}c_{1r}k_{pr}s + a_{3r}c_{1r}k_{ir}}{(a_{3r}c_{1r}k_{pr} - c_{2r})s + a_{3r}c_{1r}k_{ir}} = \frac{n_3 s + n_2}{n_1 s + n_2}, \quad (9)$$

which can be further approximated to $1/((n_1/n_2) \cdot s + 1)$. This is because the ratio of $n_3$ to $n_1$ is determined mainly by $1/V_{dcr0}$; for example, $n_3/n_1$ is less than 0.01 for the real HVDC systems [15], [21]. As shown in (8) and (9), the responsibilities to regulate $\Delta I_{dci}$ and $\Delta V_{dcr}$ can be kept distinct and assigned to separate HVDC terminals.

In addition, given the symmetry of the DC-link model, the relationship between $\Delta V_{dci}$ and $\Delta V_{dcr}$ can be represented as:

$$\Delta V_{dci} = \Delta V_{dcr} - (R + sL)(\Delta I_{dci} + sC\Delta V_{dc}) - (R + sL)\Delta I_{dci}, \quad (10)$$

where $\Delta V_{dc} = 0.5 \cdot (\Delta V_{dcr} + \Delta V_{dci})$ is the voltage variation at the

$$\frac{\Delta I_{dci}}{\Delta I_{dci\_ref}} = \frac{k_{pi}a_{3i}(b_{1i}b_{3i} + c_{2i}) \cdot s + k_{ii}a_{3i}(b_{1i}b_{3i} + c_{2i})}{\{a_{4i}b_{5i} - b_{3i}c_{1i} + k_{pi}a_{3i}(b_{1i}b_{3i} + c_{2i})\} \cdot s + k_{ii}a_{3i}(b_{1i}b_{3i} + c_{2i})} = \frac{m_3 s + m_4}{m_1 s + m_2} = \frac{m_3}{m_1} + \frac{m_4 - (m_2 m_3 / m_1)}{m_1 s + m_2} \approx \frac{1}{(m_1/m_2)s + 1} \quad (8)$$

mid-point of the DC link. Equivalently, (10) is represented as:

$$\Delta V_{dci} = \frac{1-0.5sC(R+sL)}{1+0.5sC(R+sL)}\Delta V_{dcr} - \frac{2(R+sL)}{1+0.5sC(R+sL)}\Delta I_{dci} \quad, \quad (11)$$

which can be simplified for the values of $L$ and $C$ in the real HVDC systems [15], [21] as:

$$\Delta V_{dci} \approx \Delta V_{dcr} - \frac{2R}{sT_k+1}\Delta I_{dci} \quad. \quad (12)$$

Using (12), $\Delta P_{dci}$ can be estimated as:

$$\Delta P_{dci} \approx I_{dci0}\Delta V_{dci} + V_{dci0}\Delta I_{dci} \approx I_{dci0}\left(\Delta V_{dcr} - \frac{2R}{sT_k+1}\Delta I_{dci}\right) + V_{dci0}\Delta I_{dci} \quad. \quad (13)$$

Since the variation in the capacitor charging current $\Delta I_c$ is significantly smaller than $\Delta I_{dcr}$ or $\Delta I_{dci}$ (i.e., $\Delta I_{dcr} \approx \Delta I_{dci}$), $\Delta P_{dcr}$ can also be expressed using $\Delta I_{dci}$ and $\Delta V_{dcr}$ as:

$$\Delta P_{dcr} \approx I_{dcr0}\Delta V_{dcr} + V_{dcr0}\Delta I_{dci} \quad. \quad (14)$$

It can be seen in (13) and (14) that $\Delta P_{dci}$ and $\Delta P_{dcr}$ differ only slightly under normal operating conditions: i.e., $V_{dcr0} = V_{dci0} = 1$ pu and $I_{dcr0} = I_{dci0} = 1$ pu. In other words, the control of the DC voltage and current in the proposed GFR means that the DC-link capacitor is not exploited extensively as an energy buffer, which prevents excessive operational stress on the capacitor. Using (8)–(14), the original and simplified small-signal models of the LCC HVDC system can then be developed, as shown in Fig. 2(a) and (b), respectively.

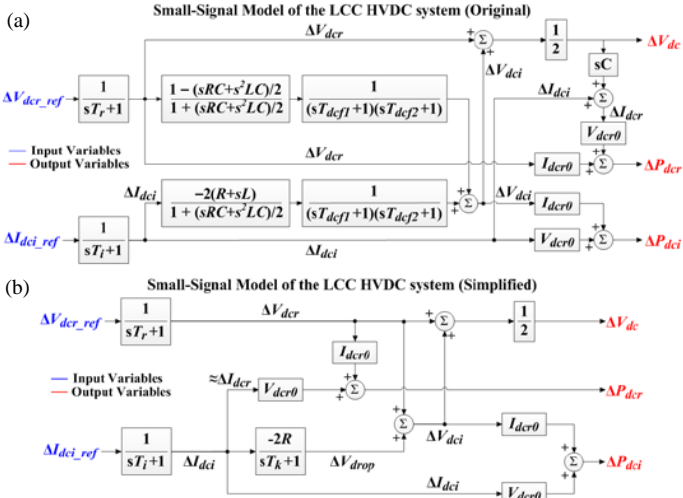

Fig. 2. (a) Original and (b) simplified small-signal models of the LCC HVDC system considering the dynamics of the DC link, HVDC converters, and inner feedback control loops.

### B. Real-time GFR Support Provided by LCC HVDC System

Fig. 3 shows a block diagram of the proposed strategy, where the LCC HVDC system supports the GFR on the rectifier- and inverter-side grids. It includes the dynamic model of the HVDC system developed in Section II-A. Moreover, for simplicity, the accumulated dynamic response of the generators on each-side grid is represented by a second-order transfer function [22] and the WF is modeled a negative load [23]. For the real-time GFR, the reference signals of the HVDC system consist of two components, corresponding to PFC and SFC, as in the case of a conventional GFR scheme for generators. The PFC is achieved using droop controllers with coefficients of $R_{i(r)}$ and $K_{i(r)}$.

Active power sharing between the HVDC converters and generators is achieved using controllers with $R_{i(r)}$, so that $\Delta f_{i(r)}$ can be stabilized in a localized manner. Similarly, using controllers with $K_{i(r)}$, $\Delta V_{dc}$ is stabilized via DC power sharing between the HVDC converters. The inertial responses are also emulated using derivative controllers to further exploit the fast dynamics of the HVDC converters.

The PFC support of an HVDC converter acts as a disturbance in the AC grid interfaced with the other converter. For example, $f_i$ decreases for an increase in $P_{li}$, activating the droop controller with $R_i$. This increases $P_{dci}$ and, consequently, $P_{dcr}$ (see (13) and (14)), leading to a decrease in $f_r$. Furthermore, the frequency deviation leads to conflict between the droop controllers. Specifically, for $\Delta f_r < 0$, the controller with $R_r$ is then activated to reduce $P_{dcr}$ and hence $P_{dci}$, whereas the controller with $R_i$ would still attempt to increase $P_{dci}$ to stabilize $f_i$. Moreover, since the reduction in $P_{dcr}$ is achieved by decreasing $V_{dcr\_ref}$, the controllers with $K_r$ and $K_i$ are activated to increase $P_{dcr}$ and decrease $P_{dci}$, respectively. In other words, DC power sharing degrades the PFC support of the controllers with $R_r$ and $R_i$. Similarly, $\Delta f_i$ resulting from $\Delta P_{lr}$ causes conflicts in the operations of the localized, droop controllers. Given the droop control characteristics, an LQG controller is designed for the coordinated SFC of $\Delta V_{dcr}$ and $\Delta I_{dci}$ for the HVDC system and $\Delta P_{gr}$ and $\Delta P_{gi}$ for the generators, as discussed in Section III, restoring $f_i$, $f_r$, and $V_{dc}$ to their nominal values.

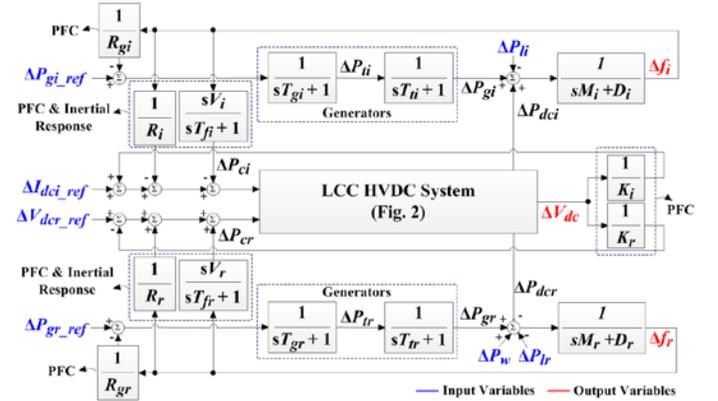

Fig. 3. Proposed GFR support by the HVDC system via inertia emulation, PFC, and SFC with variation in the load demands and WF power generation.

### III. DESIGN AND ANALYSIS OF AN LQG CONTROLLER FOR OPTIMAL SFC OF LCC HVDC SYSTEM

#### A. Design of an LQG Controller for Optimal SFC

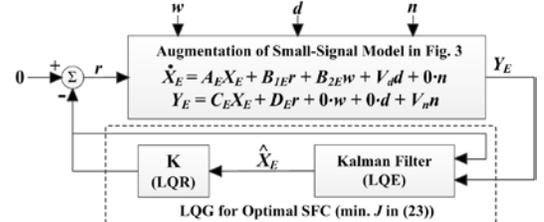

Fig. 4. An LQG controller integrated with the small-signal model (see Fig. 3) of the LCC HVDC system and conventional generators.

The small-signal model of the HVDC system and generators, shown in Fig. 3, can be represented in state space form as:

$$d\mathbf{X}(t)/dt = \mathbf{A}\cdot\mathbf{X}(t) + \mathbf{B}_r\cdot\mathbf{r}(t) + \mathbf{B}_w\cdot\mathbf{w}(t), \quad (15)$$
$$\mathbf{Y}(t) = \mathbf{C}\cdot\mathbf{X}(t),$$

with the states, inputs, and outputs arranged, respectively, as:

$$\mathbf{X}(t) = [\Delta f_i, \Delta P_{gi}, \Delta P_{ti}, \Delta P_{ci}, \Delta I_{dci},$$
$$\Delta f_r, \Delta P_{gr}, \Delta P_{tr}, \Delta P_{cr}, \Delta V_{dcr}, \Delta V_{drop}]^T \quad , \quad (16)$$

$$\mathbf{r}(t) = [\Delta P_{gi\_ref}, \Delta P_{gr\_ref}, \Delta I_{dci\_ref}, \Delta V_{dcr\_ref}]^T, \quad (17)$$

$$\mathbf{w}(t) = [\Delta P_{li}, \Delta P_{lr} - \Delta P_w]^T, \quad (18)$$

$$\mathbf{Y}(t) = \left[\Delta f_i, \Delta f_r, \Delta V_{dcr}, \Delta V_{drop}\right]^T, \quad (19)$$

where $\mathbf{A}$, $\mathbf{B_r}$, $\mathbf{B_w}$, and $\mathbf{C}$ are presented in Appendix B. Moreover, $\Delta V_{drop}$ is defined as $\Delta V_{dci} - \Delta V_{dcr}$. For optimal SFC, the state space model (15) is augmented by including the integrals of the frequency deviations and DC-link voltage variation, as:

$$\mathbf{X_E}(t) = [\int \Delta f_i \, dt, \int \Delta f_r \, dt, \int \Delta V_{dcr} \, dt, \int \Delta V_{drop} \, dt, \mathbf{X}(t)]^T, \quad (20)$$

$$\mathbf{Y_E}(t) = \left[\mathbf{Y}(t), \int \Delta f_i \, dt, \int \Delta f_r \, dt, \int \Delta V_{dcr} \, dt, \int \Delta V_{drop} \, dt\right]^T, \quad (21)$$

with the coefficient matrices augmented as:

$$\mathbf{A_E} = \begin{bmatrix} \mathbf{O} & \mathbf{A_s} \\ \mathbf{O} & \mathbf{A} \end{bmatrix}, \mathbf{B_{rE}} = \begin{bmatrix} \mathbf{O} \\ \mathbf{B_r} \end{bmatrix}, \mathbf{B_{wE}} = \begin{bmatrix} \mathbf{O} \\ \mathbf{B_w} \end{bmatrix}, \mathbf{C_E} = \begin{bmatrix} \mathbf{O} & \mathbf{C} \\ \mathbf{I} & \mathbf{O} \end{bmatrix}, \quad (22)$$

where $\mathbf{A_s}$ is specified in Appendix B. In addition, communication time delays and measurement noise are reflected in the augmented state-space model; in Fig. 4, these are modeled as input and output disturbances $\mathbf{d}(t)$ and $\mathbf{n}(t)$, respectively, using normally distributed random variables.

As shown in Fig. 4, a state feedback controller $\mathbf{r}(t) = -\mathbf{K} \cdot \mathbf{X_E}(t)$ is integrated with the augmented model to minimize the maximum variations in $f_i$, $f_r$, and $V_{dc}$, while also restoring them to the nominal values in the steady state. Considering the control efforts of the HVDC system and generators, a cost function for the feedback controller can be formulated as

$$J = \int_0^\infty \left(\mathbf{X_E}(t)^T \cdot \mathbf{Q} \cdot \mathbf{X_E}(t) + \mathbf{r}(t)^T \cdot \mathbf{R} \cdot \mathbf{r}(t)\right) dt \quad . \quad (23)$$

In (23), $\mathbf{Q}$ is a diagonal matrix with weighted coefficients, each of which is multiplied by the square of each state variable $\int \Delta f_i \, dt$, $\int \Delta f_r \, dt$, $\int \Delta V_{dcr} \, dt$, $\int \Delta V_{drop} \, dt$, $\Delta f_i$, $\Delta f_r$, $\Delta V_{dcr}$, and $\Delta V_{drop}$, in $\mathbf{X_E}(t)$. Similarly, $\mathbf{R}$ is a diagonal matrix with weighted coefficients for the squares of the input variables in $\mathbf{r}(t)$. For $\mathbf{Q}$ and $\mathbf{R}$, there is a matrix of $\mathbf{K} = \mathbf{R}^{-1} \cdot \mathbf{B_r}^T \cdot \mathbf{P}$ minimizing $J$ in (23), such that $\mathbf{P}$ is the solution to:

$$\mathbf{A_E}^T \cdot \mathbf{P} + \mathbf{P} \cdot \mathbf{A_E} + \mathbf{Q} - \mathbf{P} \cdot \mathbf{B} \cdot \mathbf{R}^{-1} \cdot \mathbf{B}^T \cdot \mathbf{P} = \mathbf{O}. \quad (24)$$

To establish the optimal $\mathbf{r}(t)$, all states in $\mathbf{X_E}(t)$ need to be measurable. As shown in Fig. 4, a Kalman filter [24] is often adopted to estimate the unknown states, using the measured states or outputs $\mathbf{Y_E}(t)$: i.e.,

$$\mathbf{r}(t) = -\mathbf{K} \cdot \mathbf{X_E}(t) \approx -\mathbf{K} \cdot \hat{\mathbf{X}}_\mathbf{E}(t), \quad (25)$$

where $\hat{\mathbf{X}}_\mathbf{E}$ is the estimate of $\mathbf{X_E}$. Given the successful performance of the Kalman filter (i.e., $\hat{\mathbf{X}}_\mathbf{E} \approx \mathbf{X_E}$), the transfer function matrix $\mathbf{V}(s)$ between $\mathbf{w}(s)$ and $\mathbf{Y_E}(s)$ of the complete state-space model, shown in Fig. 4, is obtained as:

$$\mathbf{Y_E}(s) = \mathbf{V}(s) \cdot \mathbf{w}(s) = \mathbf{C_E}\left[s\mathbf{I} - (\mathbf{A_E} - \mathbf{B_{rE}}\mathbf{K})\right]^{-1} \mathbf{B_{wE}} \cdot \mathbf{w}(s), \quad (26)$$

or, equivalently,

$$\begin{bmatrix} \Delta f_i(s) \\ \Delta f_r(s) \\ \vdots \end{bmatrix} = \begin{bmatrix} V_{11}(s) & V_{12}(s) \\ V_{21}(s) & V_{22}(s) \\ \vdots & \vdots \end{bmatrix} \begin{bmatrix} \Delta P_{li}(s) \\ \Delta P_{lr}(s) - \Delta P_w(s) \end{bmatrix} . \quad (27)$$

### B. Eigenvalue Analysis of the Proposed GFR Support by the LCC HVDC System Using the LQG Controller

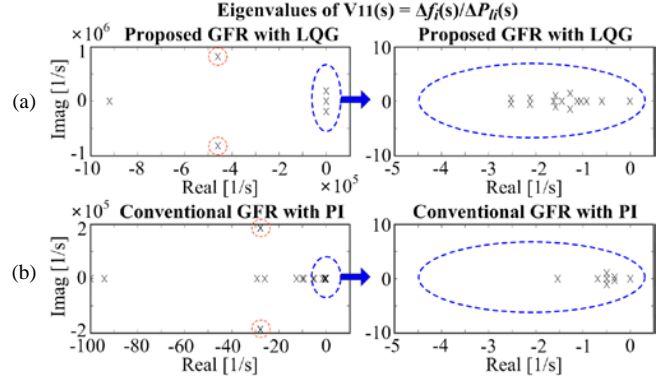

Fig. 5. Eigenvalues of $\Delta f_i/\Delta P_{li}$: (a) proposed and (b) conventional GFR.

Using (27), the eigenvalue analysis is performed focusing on the frequency deviations of the load demand variations: i.e., $\Delta P_w(s) = 0$. Fig. 5 represents the eigenvalues of $\Delta f_i(s)/\Delta P_{li}(s)$ for the proposed and conventional GFR strategies, where the LQG and proportional-integral (PI) controllers, respectively, are adopted for the SFC of the HVDC system and generators. Note that for both strategies, the eigenvalues are the same as those of $\Delta f_i(s)/\Delta P_{lr}(s)$, $\Delta f_r(s)/\Delta P_{lr}(s)$, and $\Delta f_r(s)/\Delta P_{lr}(s)$. In Fig. 5(a), all the eigenvalues are placed on the left-hand half plane, representing that the proposed strategy ensures stable operation of the HVDC system. Moreover, as shown in Fig. 5(a) and (b), the dominant poles and complex-conjugate poles with large imaginary values, marked by blue and red, respectively, are located further away from the imaginary axis for the proposed strategy than for the conventional one. This indicates that in the proposed strategy, $f_i$ can be restored to the nominal value with a faster response time and smaller oscillations.

In addition, Fig. 6 shows the root locus of $\Delta f_i(s)/\Delta P_{li}(s)$ in the proposed strategy with variation in the model parameters and controller gains of the HVDC system. In particular, the effects of the DC line parameters and droop coefficients are analyzed to verify the performance of the proposed strategy when applied to various models of the HVDC system. The DC-link parameters $L$, $R$, and $C$ increase from $0.1\times10^{-1}$ H to 1.0 H, from 0.1 Ω to 5.0 Ω, and from 10.0 μF to 200.0 μF, respectively [25], [26]. Moreover, the droop coefficients, $R_{i(r)}$ and $K_{i(r)}$, increase from 0.1 to 10.0. Note that the root locus consists of the poles obtained using different optimal values of $\mathbf{K}$ for each set of ($L$, $R$, $C$, $R_i$, $R_r$, $K_i$, $K_r$). In Fig. 6(a), as $L$ increases, the poles move towards the imaginary axis, decreasing the damping ratio and increasing the settling time of the closed-loop system (i.e., Fig. 4). Therefore, for real-time GFR support of the HVDC system, DC link and DC filters with low inductances are preferred, to move the real and complex-conjugate poles away from the imaginary axis. The variations in $R$ and $C$ have marginal effects on the eigenvalues; only the poles close to $s = -5$ move slightly toward the imaginary axis. In addition, Fig. 6(b) and (c) also show that, as $R_{i(r)}$ and $K_{i(r)}$ increase, the dominant poles on the real axis move away from the imaginary axis. This indicates that the droop controllers of the HVDC converters, discussed in Section II-B, are well coordinated in the proposed strategy, stabilizing the grid frequency successfully within a short period of time after the load disturbance. The eigenvalue analysis verifies that the proposed strategy ensures the stability of the closed-loop system for large ranges of parameters and gains, implying the wide applicability of the proposed strategy under various HVDC system and interfacing grid conditions.

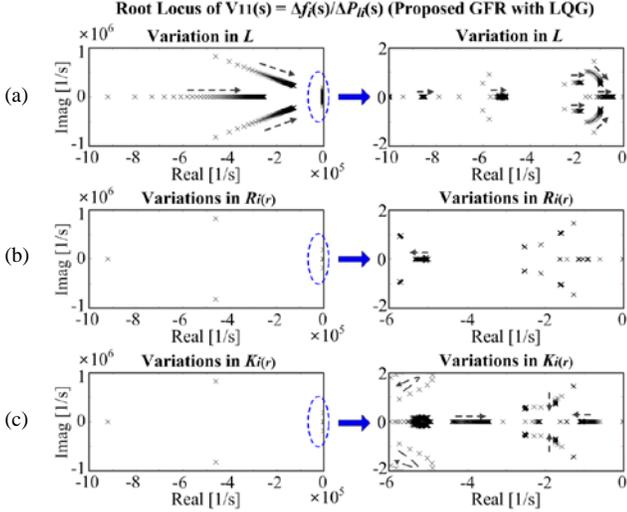

Fig. 6. Root locus of $\Delta f_i/\Delta P_{li}$ in the proposed strategy for increases in (a) $L$ from $0.1\times10^{-1}$ H to 1.0 H and in (b) $R_{i(r)}$ and (c) $K_{i(r)}$ from 0.1 to 10.0.

## IV. CASE STUDIES AND RESULTS

### A. Test System and Simulation Conditions

The Jeju-Haenam LCC HVDC system in Korea was used as a test bed to analyze the proposed and conventional control strategies. The test bed was implemented in MATLAB/SIMULINK using the averaged circuit models of the HVDC converters and the detailed models of the DC link and the inner and outer feedback controllers [15]. Table I provides the real parameters of the HVDC system and model parameters of the non-reheat turbine generators [22] in the rectifier- and inverter-side grids. It also shows the coefficients for the emulated inertial responses and droop controllers, as well as the PI controllers for the conventional strategy.

In addition, Table II shows the main features of the proposed (Case 1) and conventional (Cases 2 and 3) strategies. The comparison between Cases 1 and 2 was taken into consideration to analyze and verify the effect of the proposed LQG controller on the real-time GFR, given the operating characteristics of the droop controllers, discussed in Sections II-B and III-A. Moreover, Case 3 represents a common condition of the HVDC system [3], [4] when it is used to support only the inverter-side GFR: i.e., no feedback loops with $1/R_r$, $s\cdot V_r$, $K_r$ and $K_i$ for the rectifier-side GFR. The comparison of Case 1 with Case 3 was made to investigate and demonstrate the effectiveness of the proposed DC voltage control for improving the GFR on both rectifier and inverter sides. Furthermore, Fig. 7 shows the continuous variations in the load demands $\Delta P_{li}$ and $\Delta P_{lr}$, reflecting the scaled-up RegD signals [27] over a time period of 200 s. It also represents the intermittent power generation $\Delta P_w$ of the WF. In addition to these profiles, stepwise variations in $\Delta P_{li}$ and $\Delta P_{lr}$ were considered to evaluate the performance of the GFR support provided by the LCC HVDC system for the proposed and conventional strategies.

TABLE I. PARAMETERS FOR THE TEST BED

| Devices | Descriptions | Parameters | Values |
|---|---|---|---|
| HVDC System | Nominal DC voltages | $V_{dcr0}$, $V_{dci0}$ [kV] | 184.0, 183.5 |
| | Nominal DC currents | $I_{dcr0}$, $I_{dci0}$ [A] | 407.6 |
| | DC-link parameters | $R$ [Ω], $L$ [H], $C$ [μF] | 1.116, 0.2, 54 |
| | Converter reactance | $X_{cr}$, $X_{ci}$ [Ω] | 7.99 |
| | Converter overlap angles | $\mu_{i(r)0}$ [°] | 2.44 |
| | TR secondary voltages | $V_{lr}$, $V_{li}$ [kV] | 75.9, 82.2 |
| | TR tap ratios | $TR_{i(r)}$, | 0.9 |
| | Voltage controller gains | $k_{pr}$, $k_{ir}$ | 5.5, 20.1 |
| | Current controller gains | $k_{pi}$, $k_{ii}$ | 0.001, 10.0 |
| | Number of bridges | $N$ | 2 |
| | Time constants of HVDC converters and DC link | $T_{i(r)}$, $T_k$, $T_{fi(r)}$ | 0.02, 0.001, 0.1 |
| Generators in AC networks | Inertia and damping | $M_i$, $M_r$, $D_i$, $D_r$ | 5, 5, 1, 1 |
| | Time constants | $T_{gi}$, $T_{gr}$, $T_{ti}$, $T_{tr}$ | 0.2, 0.2, 0.5, 0.5 |
| PFC | Emulated inertia | $V_i$, $V_r$ | 5 |
| | Droop gains (HVDC) | $K_i$, $K_r$, $R_i$, $R_r$ | 0.5 |
| | Droop gains (Gen.) | $R_{gi}$, $R_{gr}$ | 0.5 |
| SFC (Conv.) | PI gains (HVDC) | $KP_r$, $KP_i$, $KI_r$, $KI_i$ | 9, 9, 6, 6 |
| | PI gains (Gen.) | $KP_r$, $KP_i$, $KI_r$, $KI_i$ | 9, 9, 6, 6 |

TABLE II. FEATURES OF THE PROPOSED AND CONVENTIONAL STRATEGIES

| HVDC Control Strategies | | $\Delta V_{dc}$ | GFR | SFC |
|---|---|---|---|---|
| Proposed | Case 1 | time-varying | both-side grids | LQG |
| Conventional | Case 2 | time-varying | both-side grids | PI |
| | Case 3 | fixed | inverter-side grid | PI |

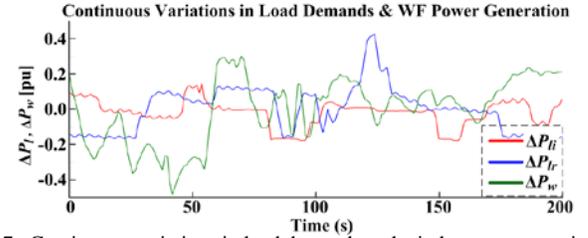

Fig. 7. Continuous variations in load demands and wind power generation.

### B. Validating the Dynamic Model of the HVDC System

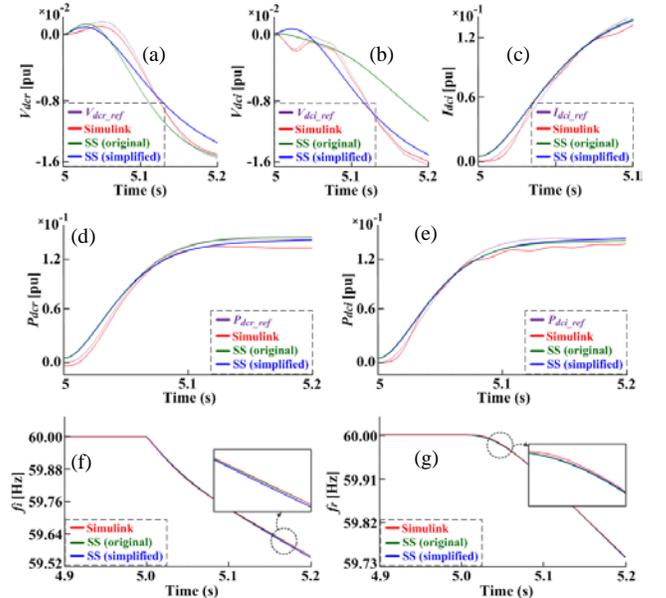

Fig. 8. Close-up plots of the step response to $\Delta P_{li}$ ($t = 5^+$ s) = 0.3 pu for the proposed strategy: (a) $V_{dcr}$, (b) $V_{dci}$, (c) $I_{dci}$, (d) $P_{dcr}$, (e) $P_{dci}$, (f) $f_i$, and (g) $f_r$.

We compared the transient responses of the original and simplified small-signal models and the comprehensive SIMULINK model of the HVDC system, discussed in Sections II-A and IV-A, respectively, to a step variation in $P_{li}$ of 0.3 pu at

$t = 5$ s. Fig. 8(a)–(c) show that the transient responses of the three HVDC system models were similar to each other for each profile of $V_{dcr}$, $V_{dci}$, and $I_{dci}$, respectively. This also led to good consistency between the simplified small-signal model and the other models for each profile of $P_{dcr}$ and $P_{dci}$ and, consequently, of $f_r$, and $f_i$, as shown in Fig. 8(d)–(g), respectively. These comparisons demonstrate the accuracy of the results obtained for the case studies using the simplified model presented in Sections IV-C, D, and E.

### C. Comparison between the Proposed and Conventional Strategies with Variation in Stepwise Load

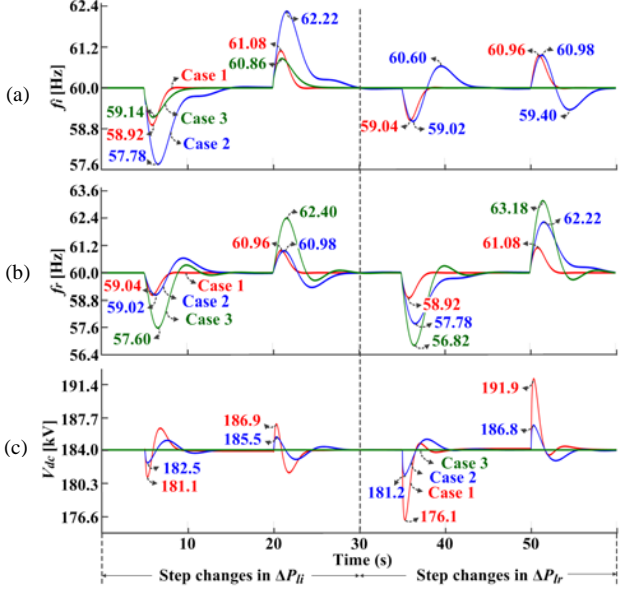

Fig. 9. Step responses of the HVDC system to $\Delta P_{li}$ ($t = 5^+$ s) = $\Delta P_{lr}$ ($t = 35^+$ s) = 0.3 pu during a 15 s for the proposed and conventional strategies: (a) $f_i$, (b) $f_r$, and (c) $V_{dc}$.

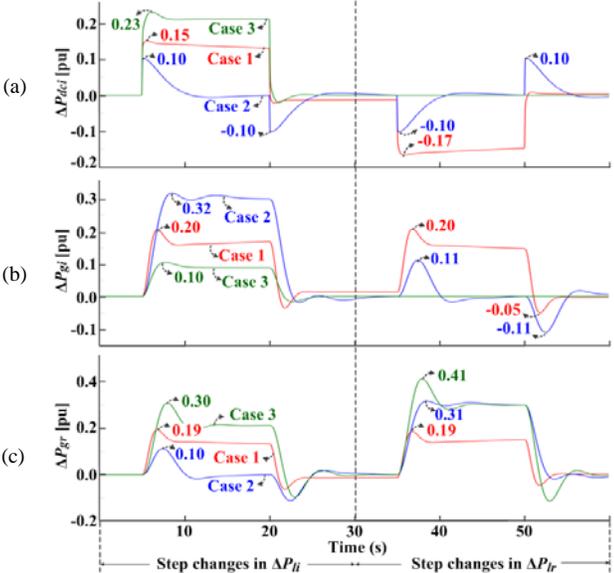

Fig. 10. Corresponding variations in the transferred and generated power: (a) $\Delta P_{dci}$, (b) $\Delta P_{gi}$, and (c) $\Delta P_{gr}$.

For the proposed and conventional strategies, Fig. 9 shows $f_i$, $f_r$, and $V_{dc}$ for the step responses of the HVDC system to $\Delta P_{li}$ and $\Delta P_{lr}$, which increased by 0.3 pu at $t = 5$ s and 35 s, respectively, over a 15 s. Specifically, the proposed strategy (i.e., Case 1) decreased the sum of the maximum deviations of $f_i$ and $f_r$ (i.e., $|\Delta f_i|_{max}+|\Delta f_r|_{max}$) by 28.8% and 21.8%, compared to the conventional strategies (i.e., Cases 2 and 3, respectively), while resulting in a maximum variation in the DC-link voltage of 4.3%. Note that although $|\Delta f_i|_{max}$ was smaller in Case 3, $|\Delta f_r|_{max}$ was considerably larger than those for Cases 1 and 2, as indicated in Table III. Moreover, in Case 1, $f_i$ and $f_r$ were restored back to the nominal value (i.e., 60 Hz) more rapidly, and with smaller overshooting than in Cases 2 and 3.

Fig. 10 represents the corresponding profiles of $\Delta P_{dci}$, $\Delta P_{gi}$, and $\Delta P_{gr}$. Note that $\Delta P_{dcr}$ was slightly different to $\Delta P_{dci}$ due to the power loss along the DC link, as discussed in Section II-A. In Case 1, both $\Delta P_{gi}$ and $\Delta P_{gr}$ increased by approximately 0.15 pu (or, more accurately, 0.16 pu and 0.14 pu, respectively) when $\Delta P_{li}$ or $\Delta P_{lr}$ increased by 0.30 pu, given the identical model parameters and controller gains of the rectifier- and inverter-side generators. In other words, the LQG controller enabled the load demand variation to be shared between the generators in both-side grids, reducing grid frequency deviations and preventing excessive operational stress on the generators. Optimal control of the DC-link voltage and current was also achieved to transfer the increased power generation in both the transient and steady state. In contrast, the PI controllers in Case 2 led the generators to compensate for the load variations, which mainly occurred on the same side of the HVDC system. For $\Delta P_{li}$ ($t = 5^+$ s) = 0.30 pu, $\Delta P_{gi}$ increased by approximately 0.30 pu, while $\Delta P_{gr}$ initially increased by 0.10 pu and then remained close to zero after the transient state time (i.e., $t > 16$ s); this is similar to the case of $\Delta P_{lr}$ ($t = 35^+$ s) = 0.30 pu. Therefore, the HVDC system could only marginally contribute to the rectifier-side GFR, and only during the period of the transient state, resulting in larger deviations and overshooting of $f_i$ and $f_r$, as shown in Fig. 9. In Case 3, $\Delta P_{li}$ ($t = 5^+$ s) = 0.3 pu caused larger variations in $P_{gr}$ and hence $f_r$ than in $P_{gi}$ and $f_i$, which is particularly problematic when the rectifier-side grid includes critical loads and generators with limited capacities. For all cases, $\Delta P_{dci}$ and $\Delta P_{dcr}$ changed faster than $\Delta P_{gi}$ and $\Delta P_{gr}$. This verifies the effectiveness of the inertia emulation and droop control approaches in improving the GFR by better exploiting the fast dynamics of the HVDC converters.

TABLE III. MAXIMUM VARIATIONS IN FREQUENCY, TRANSFERRED POWER, AND GENERATED POWER IN THE STEP RESPONSE TEST OF HVDC SYSTEM

| Max. variations | Case 1 Individual | Case 1 Total | Case 2 Individual | Case 2 Total | Case 3 Individual | Case 3 Total |
|---|---|---|---|---|---|---|
| $|\Delta f_i|_{max}$ [Hz] | 1.08 | 3.16 | 2.22 | 4.44 | 0.86 | 4.04 |
| $|\Delta f_r|_{max}$ [Hz] | 1.08 | | 2.22 | | 3.18 | |
| $|\Delta P_{dci}|_{max}$ [pu] | 0.17 | 0.34 | 0.10 | 0.20 | 0.23 | 0.46 |
| $|\Delta P_{dcr}|_{max}$ [pu] | 0.17 | | 0.10 | | 0.23 | |
| $|\Delta P_{gi}|_{max}$ [pu] | 0.20 | 0.39 | 0.32 | 0.63 | 0.10 | 0.51 |
| $|\Delta P_{gr}|_{max}$ [pu] | 0.19 | | 0.31 | | 0.41 | |

### D. Comparisons between the Proposed and Conventional Strategies for Continuous Load Variations

Additional case studies were carried with continuous variation of $\Delta P_{li}$, $\Delta P_{lr}$, and $\Delta P_w$, shown in Fig. 7. Fig. 11 shows the corresponding profiles of $f_i$, $f_r$, and $V_{dc}$. Table IV also lists the RMS variations in $\Delta f$ and $\Delta P_g$, which were estimated as:

$$\Delta f_{rms} = \sqrt{\frac{1}{M}\sum_{m=1}^{M}\Delta f_m^2} \quad \text{and} \quad \Delta P_{g,rms} = \sqrt{\frac{1}{M}\sum_{m=1}^{M}\Delta P_{g,m}^2}, \quad (28)$$

where $m$ is the index of the measurement sample and $M$ is the total number of samples. In Case 1, the sum of the RMS variations in $f_i$ and $f_r$ decreased by 57.3 % and 53.1%, compared to Cases 2 and 3, respectively, demonstrating the effectiveness of the proposed HVDC system control in improving the GFR in both rectifier- and inverter-side networks. Moreover, $\Delta P_{g,rms}$ in Case 1 was 23.3% and 8.0% smaller than those in Cases 2 and 3, respectively, because the total variation in the load demand and WF power generation was shared among all generators on both sides. This implies that the proposed control strategy can effectively mitigate the operational requirements (e.g., spinning reserve capacity) of the generators, so that the cost due to the increase in power transferred via the HVDC system can be compensated for by savings in the generator operating costs resulting from the increased flexibility in the real-time GFR.

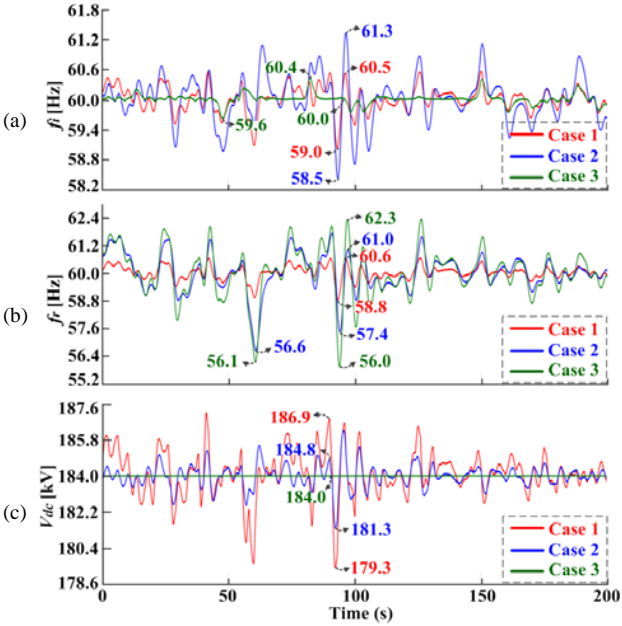

Fig. 11. Responses of the HVDC system to continuous variations in $\Delta P_{li}$, $\Delta P_{lr}$, and $\Delta P_w$ for the proposed and conventional strategies: (a) $f_i$, (b) $f_r$, and (c) $V_{dc}$.

TABLE IV. RMS VARIATIONS IN FREQUENCY AND POWER GENERATION IN THE CONTINUOUS RESPONSE TEST OF HVDC SYSTEM

| RMS variations | Case 1 Individual | Case 1 Total | Case 2 Individual | Case 2 Total | Case 3 Individual | Case 3 Total |
|---|---|---|---|---|---|---|
| $\Delta f_{i,rms}$ [Hz] | 0.25 | 0.53 | 0.44 | 1.24 | 0.10 | 1.13 |
| $\Delta f_{r,rms}$ [Hz] | 0.28 |  | 0.80 |  | 1.03 |  |
| $\Delta P_{gi,rms}$ [pu] | 0.11 | 0.23 | 0.08 | 0.30 | 0.02 | 0.25 |
| $\Delta P_{gr,rms}$ [pu] | 0.12 |  | 0.22 |  | 0.23 |  |

*E. Performance of the Proposed Strategy under Various Test Conditions*

The case studies discussed in Section IV-C were repeated to analyze the performance of the proposed strategy under various HVDC system conditions, particularly with respect to **Q** in (23). Table V lists the results of the case studies regarding the step responses of the HVDC system, when the weighting coefficients in **Q** were increased by 10 times, compared to the original values used in Section IV-C. From the comparison of Table V with those in Fig. 9 and Table III, it can be seen that the proposed strategy decreased both the frequency and DC-link voltage variations. In other words, the transient stability of both the HVDC system and the interfacing grids was enhanced, although the DC power loss increased by 3.5% due to an increase in the DC current. Furthermore, when we increased the weighting coefficients in **Q** assigned to the states $\Delta f_i$ and $\Delta f_r$ in $\mathbf{X}_E(t)$, the maximum frequency deviation decreased gradually. This implies that the proposed strategy enables the HVDC system to support the rectifier-side GFR more intensively than the inverter-side GFR, or vice versa, according to, for example, load composition and the installed capacity of dispatchable generators and wind turbines; this is left for future work.

The case studies in Section IV-D were also repeated to analyze the effects of the inertia emulation and droop control conditions. The LQG controller introduced in Section III successfully reduced the frequency deviations under all conditions via optimal coordination of the HVDC converters and generators; Table VI lists the numerical results.

TABLE V. VARIATIONS IN GRID FREQUENCY, DC VOLTAGE, TRANSFERRED POWER, AND GENERATED POWER FOR **Q** THAT INCREASED BY 10 TIMES.

| Frequency [Hz] | | | DC-link voltage [kV] | |
|---|---|---|---|---|
| $|\Delta f_i|_{max}$ | $|\Delta f_r|_{max}$ | Total | $|\Delta V_{dc}|_{max}$ for $\Delta P_{li}$ | $|\Delta V_{dc}|_{max}$ for $\Delta P_{lr}$ |
| 0.96 | 0.96 | 1.92 | 0.7 | 2.0 |
| Transferred power [pu] | | | Generated power [pu] | |
| $|\Delta P_{dci}|_{max}$ | $|\Delta P_{dcr}|_{max}$ | Total | $|\Delta P_{gi}|_{max}$ | $|\Delta P_{gr}|_{max}$ | Total |
| 0.16 | 0.16 | 0.32 | 0.18 | 0.16 | 0.34 |

TABLE VI. RMS VARIATIONS IN THE FREQUENCY AND GENERATED POWER UNDER THE INERTIAL RESPONSE AND DROOP CONTROL CONDITIONS

| | RMS variations | Case 1 Individual | Case 1 Total | Case 2 Individual | Case 2 Total | Case 3 Individual | Case 3 Total |
|---|---|---|---|---|---|---|---|
| a | $\Delta f_{i\,rms}$ [Hz] | 0.24 | 0.52 | 0.53 | 1.47 | 0.11 | 1.14 |
|  | $\Delta f_{r\,rms}$ [Hz] | 0.28 |  | 0.94 |  | 1.03 |  |
|  | $\Delta P_{gi\,rms}$ [pu] | 0.11 | 0.23 | 0.09 | 0.32 | 0.02 | 0.25 |
|  | $\Delta P_{gr\,rms}$ [pu] | 0.12 |  | 0.23 |  | 0.23 |  |
| b | $\Delta f_{i\,rms}$ [Hz] | 0.25 | 0.53 | 0.37 | 1.38 | 0.10 | 1.13 |
|  | $\Delta f_{r\,rms}$ [Hz] | 0.28 |  | 1.01 |  | 1.03 |  |
|  | $\Delta P_{gi\,rms}$ [pu] | 0.11 | 0.23 | 0.08 | 0.31 | 0.02 | 0.25 |
|  | $\Delta P_{gr\,rms}$ [pu] | 0.12 |  | 0.23 |  | 0.23 |  |

a: no droop control, b: no inertia emulation and no droop control

## V. CONCLUSIONS

This paper has proposed a new control strategy for an LCC HVDC system, in which the DC-link voltage and current were optimally regulated to improve the real-time GFR in both the rectifier- and inverter-side networks. A new small-signal model of an LCC HVDC system was developed, verified via comparison with a comprehensive model, and integrated with feedback loops for inertia emulation and droop control. An LQG controller including a Kalman filter was also designed for optimal SFC while mitigating conflict between the droop controllers of the HVDC converters. An eigenvalue analysis was then conducted, focusing on the effects of the DC-link parameters and droop coefficients on the performance of the proposed strategy. Simulation case studies were also carried out using the Jeju-Haenam HVDC system as a test bed, where the proposed strategy decreased the maximum frequency variation by 28.8% and the RMS variation by 57.3% for stepwise and continuous load demand variations, respectively, compared to the conventional one using the PI controllers. The proposed strategy enabled the total load demand variation to be shared between the generators in both-side grids, reducing the RMS variation in the power generation by 23.3%. The proposed strategy was effective under various conditions of the weighting coefficients $\mathbf{Q}$ and the inertia emulation and droop control approaches; for higher $\mathbf{Q}$, the proposed strategy decreased variation in both the frequency and DC-link voltage.

## APPENDIX

### A. Coefficients for the Dynamic Model of HVDC System

The coefficients used in (1)–(9) are defined as

$$a_{1x} = 3\sqrt{2}NV_{lx}(\pi TR_x)^{-1}, \quad a_{2x} = 3X_{cx}N\pi^{-1}, \quad a_{3x} = H_{dcx\_0}^{-1}, \quad (A1)$$

$$a_{4x} = \sqrt{2}X_{cx}/(V_{lx}/TR_x), \quad a_{5x} = 3N(V_{lx})^2/4\pi X_{cx}(TR_x)^2,$$

$$b_{1x} = -a_{1x}\sin\beta_0, \quad b_{2x} = -\sin\beta_0 + \sin(\beta_0 + \mu_{x0}), \quad (A2)$$

$$b_{3x} = \sin(\beta_0 + \mu_{x0}), \quad b_{4x} = a_{5x}\{-2\sin(2\beta_0) + 2\sin 2(\beta_0 + \mu_{x0})\},$$

$$b_{5x} = a_{5x}2\sin 2(\beta_0 + \mu_{x0}),$$

$$c_{1i} = V_{dci0} - a_{2i}I_{dci0}, \quad c_{2i} = b_{2i}b_{5i} - b_{3i}b_{4i},$$

$$c_{1r} = (b_{1r}b_{3r}V_{dcr0} - a_{4r}b_{1r}b_{5r} - a_{2r}b_{3r}b_{4r} + a_{2r}b_{2r}b_{5r}), \quad (A3)$$

$$c_{2r} = (b_{3r}V_{dcr0} - a_{4r}b_{5r} - a_{2r}b_{3r}I_{dcr0}),$$

where the letter set of ($x$, $\beta$, $H$) corresponds to ($r$, $\alpha$, $V$) and ($i$, $\gamma$, $I$) for the rectifier and inverter, respectively.

### B. Coefficients for the State-Space Model of HVDC System

For the proposed strategy, the coefficient matrices for the complete state-space model of the LCC HVDC system, interfaced with AC networks (see Fig. 4), are represented as:

$$\mathbf{A_s} = \begin{bmatrix} 1 & [\mathbf{O}]_{1\times 5} & [\mathbf{O}]_{1\times 5} \\ [\mathbf{O}]_{1\times 5} & 1 & [\mathbf{O}]_{1\times 5} \\ [\mathbf{O}]_{1\times 9} & 1 & 0 \\ [\mathbf{O}]_{1\times 5} & [\mathbf{O}]_{1\times 5} & 1 \end{bmatrix}, \quad \mathbf{A} = \begin{bmatrix} [\mathbf{A_1}] & [\mathbf{O}]_{1\times 5} & [\mathbf{O}]_{2\times 1} \\ [\mathbf{O}]_{1\times 5} & [\mathbf{A_1}] & 1 \\ [\mathbf{O}]_{14\times 3} & [\mathbf{A_2}] & [\mathbf{O}]_{5\times 5} & [\mathbf{A_4}] \\ [\mathbf{O}]_{5\times 4} & [\mathbf{A_3}] & [\mathbf{A_5}] \\ [\mathbf{A_6}] & [\mathbf{O}]_{1\times 5} & -1/T_k \end{bmatrix}, \quad (A4)$$

where

$$[\mathbf{A_1}] = \begin{bmatrix} 1 & [\mathbf{O}]_{1\times 4} \\ 0 & [\mathbf{O}]_{1\times 4} \end{bmatrix},$$

$$[\mathbf{A_4}] = \begin{bmatrix} [\mathbf{O}]_{4\times 1} \\ 1/k_i T_i V_{dci0} \end{bmatrix}, \quad [\mathbf{A_2}] = \begin{bmatrix} -D_i/M_i & 0 & 1/M_i & 0 & 1/M_i \\ -1/R_{gi}T_{gi} & -1/T_{gi} & 0 & 0 & 0 \\ 0 & 1/T_{ti} & -1/T_{ti} & 0 & 0 \\ -V_i D_i/M_i T_{fi} & 0 & V_i/M_i T_{fi} & -1/T_{fi} & V_i/M_i T_{fi} \\ -1/R_i T_i V_{dci0} & 0 & 0 & -1/T_i V_{dci0} & -1/T_i \end{bmatrix},$$

$$[\mathbf{A_5}] = \begin{bmatrix} [\mathbf{O}]_{4\times 1} \\ -1/k_r T_r I_{dcr0} \end{bmatrix}, \quad [\mathbf{A_6}] = \begin{bmatrix} [\mathbf{O}]_{1\times 4} & -2R/T_k \end{bmatrix},$$

$$[\mathbf{A_3}] = \begin{bmatrix} -1/M_r & -D_r/M_r & 0 & -1/M_r & 0 & 0 \\ 0 & -1/R_{gr}T_{gr} & -1/T_{gr} & 0 & 0 & 0 \\ 0 & 0 & 1/T_{tr} & -1/T_{tr} & 0 & 0 \\ -V_r/M_r T_{fr} & -V_r D_r/M_r T_{fr} & 0 & V_r/M_r T_{fr} & -1/T_{fr} & 0 \\ 0 & 1/R_r T_r I_{dcr0} & 0 & 0 & 1/T_r I_{dcr0} & -1/T_r \end{bmatrix},$$

$$\mathbf{B_E} = \mathbf{B_{1E}} + \mathbf{B_{2E}} = \begin{bmatrix} [\mathbf{O}]_{3\times 6}^T & [\mathbf{B_1}]^T & [\mathbf{B_2}]^T & [\mathbf{O}]_{1\times 6}^T \end{bmatrix}^T, \quad (A5)$$

where

$$[\mathbf{B_1}] = \begin{bmatrix} [\mathbf{O}]_{1\times 4} & -1/M_i & 0 \\ 1/T_{gi} & 0 & [\mathbf{O}]_{1\times 4} \\ & [\mathbf{O}]_{1\times 6} & \\ [\mathbf{O}]_{1\times 4} & -V_i/M_i T_{fi} & 0 \\ [\mathbf{O}]_{1\times 2} & 1/T_i V_{dcr0} & [\mathbf{O}]_{1\times 3} \end{bmatrix}, \quad [\mathbf{B_2}] = \begin{bmatrix} [\mathbf{O}]_{1\times 4} & 0 & -1/M_r \\ 0 & 1/T_{gr} & [\mathbf{O}]_{1\times 4} \\ & [\mathbf{O}]_{1\times 6} & \\ [\mathbf{O}]_{1\times 4} & 0 & -V_r/M_r T_{fr} \\ [\mathbf{O}]_{1\times 3} & 1/T_r I_{dcr0} & [\mathbf{O}]_{1\times 3} \end{bmatrix},$$

and

$$\mathbf{C_E} = \begin{bmatrix} [\mathbf{O}]_{1\times 3} & 1 & [\mathbf{O}]_{1\times 10} \\ [\mathbf{O}]_{1\times 8} & 1 & [\mathbf{O}]_{1\times 5} \\ [\mathbf{O}]_{1\times 13} & 1 \\ [\mathbf{C_1}] & [\mathbf{O}]_{3\times 10} \end{bmatrix}, \quad [\mathbf{C_1}] = \mathbf{I}_{3\times 3}, \quad \mathbf{D_E} = [\mathbf{O}]_{6\times 4}. \quad (A6)$$